\documentclass{elsart}
\usepackage {epsfig,amsmath}
\begin{document}
\begin{frontmatter}

\title{ High Altitude test of RPCs for the Argo YBJ experiment}
\collab{The ARGO-YBJ Collaboration}
\author[a]{C. Bacci}, \author[b]{K.Z. Bao}, \author[c]{F. Barone},
\author[c]{B. Bartoli}, \author[d]{P. Bernardini},
\author[c]{R. Buonomo}, \author[a]{S. Bussino}, \author[c]{E. Calloni},
\author[e]{B.Y. Cao}, \author[f]{R. Cardarelli}, \author[c]{S. Catalanotti},
\author[f]{A. Cavaliere}, \author[d]{F. Cesaroni}, \author[a]{P. Creti},
\author[g]{Danzengluobu}, \author[c]{B. D'Ettorre Piazzoli},
\author[a]{M. De Vincenzi}, \author[c]{T. Di Girolamo},
\author[c]{G. Di Sciascio},
\author[h]{Z.Y. Feng}, \author[e]{Y. Fu}, \author[i]{X.Y. Gao},
\author[i]{Q.X. Geng}, \author[g]{H.W. Guo}, \author[j]{H.H. He},
\author[e]{M. He}, \author[h]{Q. Huang}, \author[c]{M. Iacovacci},
\author[a]{N. Iucci}, \author[h]{H.Y. Jai}, \author[e]{F.M. Kong},
\author[j]{H.H. Kuang}, \author[g]{Labaciren}, \author[b]{B. Li},
\author[e]{J.Y. Li}, \author[i]{Z.Q. Liu}, \author[j]{H. Lu},
\author[j]{X.H. Ma},
\author[d]{G. Mancarella}, \author[k]{S.M. Mari}, \author[d]{G. Marsella},
\author[d]{D. Martello}, \author[g]{D.M Mei}, \author[g]{X.R. Meng},
\author[c]{L. Milano}, \author[f]{A. Morselli}, \author[i]{J. Mu},
\author[d]{M. Panareo},
\author[a]{M. Parisi}, \author[a]{G. Pellizzoni}, \author[j]{Z.R. Peng},
\author[d]{C. Pinto},
\author[a]{P. Pistilli},\author[f]{E. Reali} \author[f]{R. Santonico}\footnote{E-mail address
santonico@roma2.infn.it; fax number + 39 06 2023507.},
\author[l]{G. Severino},
\author[j]{P.R. Shen},
\author[a]{C. Stanescu}, \author[j]{J. Su},
\author[b]{L.R. Sun}, \author[b]{S.C. Sun}, \author[d]{A. Surdo},
\author[j]{Y.H. Tan}, \author[l]{S. Vernetto},
\author[e]{C.R. Wang}, \author[j]{H. Wang}, \author[j]{H.Y. Wang},
\author[b]{Y.N. Wei}, \author[j]{H.T. Yang}, \author[b]{Q.K. Yao},
\author[h]{G.C. Yu}, \author[b]{X.D. Yue}, \author[g]{A.F. Yuan},
\author[j]{H.M. Zhang}, \author[j]{J.L. Zhang}, \author[e]{N.J. Zhang},
\author[i]{T.J. Zhang}, \author[e]{X.Y. Zhang}, \author[g]{Zhaxisangzhu},
\author[g]{Zhaxiciren}, \author[j]{Q.Q. Zhu}

\newpage
\address[a]{INFN and Dipartimento di Fisica dell'Universit\`a di Roma Tre,
Italy}
\address[b]{Zhenghou University, Henan, China}
\address[c]{INFN and Dipartimento di Fisica dell'Universit\`a di Napoli, Italy}
\address[d]{INFN and Dipartimento di Fisica dell'Universit\`a di Lecce, Italy}
\address[e]{Shangdong University, Jinan, China}
\address[f]{INFN and Dipartimento di Fisica dell'Universit\`a di Roma
"Tor Vergata", Italy}
\address[g]{Tibet University, Lhasa, China}
\address[h]{South West Jiaotong University, Chengdu, China}
\address[i]{Yunnan University, Kunming, China}
\address[j]{IHEP, Beijing, China}
\address[k]{Universit\'a della Basilicata, Potenza, Italy}
\address[l]{Istituto di Cosmogeofisica del CNR and INFN, Torino, Italy}

\begin{abstract}
A 50 $ m^2 $ RPC carpet was operated at the YanBaJin Cosmic Ray
Laboratory (Tibet) located 4300 m a.s.l. The performance of RPCs
in detecting Extensive Air Showers was studied. Efficiency and
time resolution measurements at the pressure and temperature
conditions typical of high mountain laboratories, are reported.
\end{abstract}
\end{frontmatter}

\section{\bf{Introduction}}
The aim of the ARGO-YBJ experiment is the study of cosmic rays,
mainly $\gamma$-radiation, in an energy range down to about 100
GeV, by detecting small size air showers with a ground detector.
This very low energy threshold, which is below the upper limit of
the next generation satellite experiments, is achieved in two
ways:

1-By operating the experiment at very high altitude to better
approach the level where low energy air showers reach their
maximum development. The choice of the YangBaJing (YBJ) Cosmic Ray
Laboratory (Tibet, China, $30.11^{\circ}$ N, $90.53^{\circ}$ E.),
4300 m a.s.l, was found to be very appropriate.

2-By utilizing a full coverage detector to maximize the number of detected
  particles for a small size shower.
\par
The choice of the detector is subject to the following
requirements.
 The search for point sources requires the accurate
reconstruction of the shower parameters, mainly the direction of
the primary particle, in order to suppress the isotropic
background. This can be obtained by a diffuse sampling on the
arrival times of the shower front particles with nanosecond
accuracy. Moreover the full coverage concept requires an extremely
large active detector area which is only achievable with a very
reliable and low cost detector. Robustness is a further important
requirement for a detector to be operated far away from the
facilities available in ordinary laboratories. The use of
Resistive Plate Chambers (RPCs) has been envisaged to meet these
requirements. Indeed, RPCs offer noticeable advantages owing to
low cost, large active area, excellent time resolution and
possibility of an easy integration in large systems.
\begin{figure}
  \begin{center}
    \includegraphics[width=14cm]{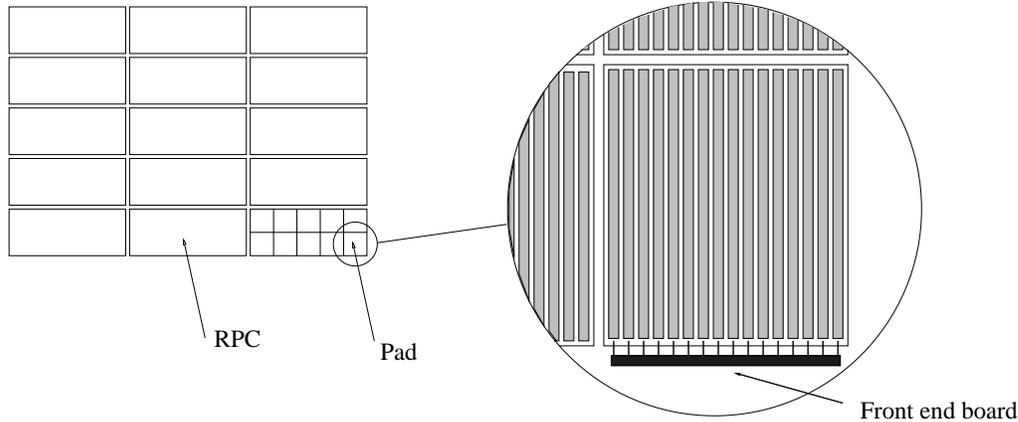}
    \caption{Layout of the CLUSTER prototype
which has been tested. Each RPC is subdivided in 10 Pads. The
details of the Pad are also shown.}
    \label{1}
  \end{center}
\end{figure}
\par
The ARGO-YBJ detector consists of a single RPC layer of $\sim
5000$ $m^2$ and about $95 \%$ coverage, surrounded by a ring of
sampling stations which recover edge effects and increase the
sampling area for showers initiated by $>5$ TeV primaries.

The trigger and the DAQ systems are built following a two level
architecture. The signals of a set of 12 contiguous RPCs, referred to as CLUSTER in the
following, are managed by a Local Station.
The information from each Local Station is collected and elaborated in the
Central Station. According to this logic a CLUSTER represents the basic detection unit.

A CLUSTER prototype of 15 RPCs, shown in fig 1, has been put in
operation in the YBJ Laboratory in order to check both the
performance of RPCs operated in a high altitude laboratory and
their capability of imaging a small portion of the shower front.

In this paper the results concerning the performance of 2 mm gap, bakelite RPC
detectors operated in streamer mode at an atmospheric depth of 606 $g/cm^2$
are described. Data collected by the carpet and results from their analysis
will be presented elsewhere.

\section{ \bf {The experimental set up}}

\begin{figure}
  \begin{center}
    \leavevmode
    \includegraphics[width=10cm]{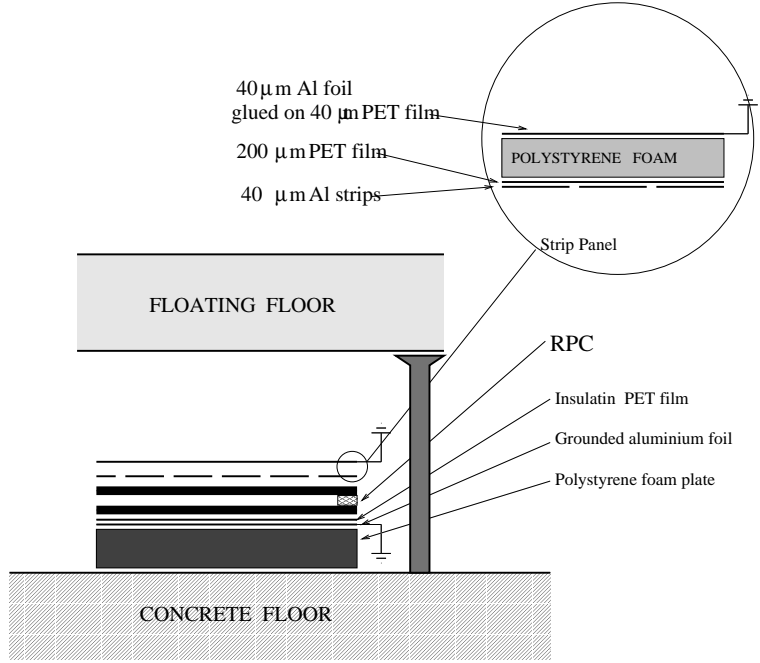}
    \caption{Cross section of the detector with details of the strip panel}
    \label{2}
  \end{center}
\end{figure}

The detector, consisting of a single gap RPC layer, is installed
inside a dedicated building at the YBJ laboratory. The set up,
shown in fig. 1, is an array of 3x5 chambers of area $280 \times
112 cm^{2}$ each, laying on the building floor and covering a
total area of $ 8.5 \times 6.0 m^{2}$. The active area of $46.2
m^{2}$, accounting for a dead area due to a $7 mm$ frame closing
the chamber edge, corresponds to a $90.6 \%$ coverage. The RPCs,
of 2 mm gas gap, are built with bakelite electrode plates of
volume resistivity in the  range ($0.5\div 1$)  $ 10^{12} \Omega
\cdot cm$, according to the standard scheme reported in \cite{1}.
The RPC signals are picked up by means of aluminum strips 3.3 cm
wide and 56 cm long which are glued on a 0.2 mm thick film of
plastic material (PET \footnote{Poly-Ethylen-Tereftalate}) used as
a robust support which allows to work out the strips by milling a
full aluminum layer. The strips are embodied in a panel,
consisting of a 4 mm thick polystyrene foam sheet sandwitched
between the PET film and an aluminum foil used as a ground
reference electrode.
  The detector cross section is given in fig. \ref{2}.
  \par
A rigid polystyrene foam plate is used to avoid the direct contact
of the RPCs with the concrete floor.  The strip panel lays on top
of the detector with the strips oriented in the direction of the
detector short side as shown in fig. \ref{1} . At the edge of the
detector the strips are connected to the front end electronics and
terminated with 50 $\Omega$ resistors. The opposite end of the
strips, at the center of the detector, is not terminated. The RPC
bottom electrode plate is connected to a negative high voltage so
that the strips, facing the grounded plate, pick up a negative
signal. A grounded aluminum foil (see fig. \ref{2}) is used to
shield the bottom face of the RPC and an extra PET foil, on top of
the aluminum, is used as a further high voltage insulator.
\par
The
front end electronics that has been used in the present test is
not the one envisaged for the final experiment, which will be
described elsewere, but is an already existing 16 channel circuit
\cite{2} developed for RPCs working in streamer mode. The circuit
contains 16 discriminators with about 50 mV voltage threshold and
gives the following output signals:

\begin{itemize}
  \item The Fast OR of the 16 discriminators with the same
input-to-output delay (10 ns) for all the channels, which is used
for time measurements and trigger purposes in the present test.

  \item Serial read out of each channel that could be used for a strip by
   strip read out. This possibility however is beyond the purposes of
   the present test.
\end{itemize}

The circuit is mounted on a $50 \times 15 cm^{2}$ G10 board which is fixed
on top of the strip panel near to the edge of the detector as
shown in fig. \ref{1}. The length of the board is approximately tuned
with the width of 16 strips so that very short wires (a few cm)
can be used for connecting each strip to the corresponding input
electrode on the board.
\par
The 16 strips connected to the same front end board are logically
organized in a PAD of $56 \times 56 cm^{2}$ area. Each RPC is
therefore subdivided in 10 PADs which work like independent
functional units. The PADs are the basic elements which define the
space-time pattern of the shower; they give indeed the position
and the time of each detected hit. The fast OR signals of all 150
pads are sent through coaxial cables of the same length to the
carpet central trigger and read out electronics.
\par
The trigger logics allows to select events with a pad multiplicity
in excess of a given threshold. At any trigger occurrence the
times of all the pads are read out by means of multihit TDCs of 1
ns time bin, operated in common STOP mode. Each TDC has 32 input
channels and can store up to 16 events per channel. The multiple
hit operation is particularly important in detecting the core of
high energy showers where several particles can fall on the same
pad in a time interval of hundreds of nanoseconds. The trigger
signal is used as the common STOP signal. For each event the
trigger multiplicity, the set of all pads which produced the
trigger and the times of all pads of the carpet are recorded.
\par
As the carpet consists just of a single layer detector, a direct
measurement of the detection efficiency and time resolution
requires the use of an auxiliary "telescope" which can clearly
define a cosmic ray impinging on it. The set up was therefore
completed with a small telescope consisting of  3 RPCs of $50
\times 50 cm^{2}$ area with 16 pick up strips 3 cm wide connected
to front end electronics boards similar to the ones used in the
carpet. The 3 RPCs were overlapped one on the other and the triple
coincidence of their fast OR signals was used to define a cosmic
ray crossing the telescope.
\par
The gas system consisted of a central mixing station using three
mass flowmeters that measured the gas composition with the require
accuracy, better than 1$\%$ for all the components, and 5 parallel
gas lines each feeding 3 RPCs in series. The gas sharing among the
5 input lines was equalized using identical high impedance
capillar pipes in series with each line and the regular gas flow
was monitored by bubblers put at the exit of each line. An open
gas circuit was used, as only a modest amount of gas, about 15 l/h
corresponding to 4 volume changes per day, was needed during about
2 months of carpet operation. Three gas components were used:
Argon, iso-Butane $C_{4}H_{10}$ and TetraFluoroEthane
$C_{2}H_{2}F_{4}$ that will be indicated in the following as Ar,
i-But and TFE respectively.
\par
The High Voltage system consisted of five 10 kV supplies each one
feeding 3 RPCs in parallel. The operating voltage was settable to
the wanted value within 10 V accuracy and the operating current
was monitored with a 1 $\mu$A sensitivity instrument. A further
two channel HV supply with 10 nA sensitivity current monitor was
used to feed the auxiliary telescope.

\section { \bf{Data taking and experimental results}}

The peculiar working conditions of the mountain YBJ laboratory are
not only a very low average pressure of about 600 mbar,
corresponding to an atmospheric vertical depth of 606 $g$
$cm^{-2}$, but also a temperature that could be particularly low
in winter even inside the laboratory.
\par
The measurements described in this paper were performed in the
2$^{nd}$ half of February 1998 with an external temperature
ranging between -20 and -5 $^{o}C$ and in the 1st half of May when
the temperature was in the range -5 +15 $^{o}C$. The internal
temperature was kept, by using some heaters, between +4 and +8
$^{o}C$ in the first run and around 16-18 $^{o}C$ in the second.
The laboratory temperature and pressure were monitored during all
data taking.
\par
The RPCs of the test carpet were operated in streamer mode \cite
{3} as foreseen for the final experiment. This mode delivers \cite
{4} large amplitude saturated signals, and is less sensitive than
the avalanche or proportional mode \cite {5} to electromagnetic
noise, to changes in the environment conditions and to mechanical
deformations of the detector. On the other hand the larger rate
capability achievable in avalanche mode \cite {6} is not needed in
a cosmic ray experiment.
\par
The first task to be carried out was the optimization of the gas
mixture and the search for the detector operating point in the YBJ
laboratory conditions. This was accomplished by means of the
auxiliary telescope, before the start up of the carpet test. The
efficiency of the RPC in the central position of the telescope
(RPC2 in the following) was measured as the ratio of the number of
triple coincidence events to the number of double coincidences of
the other two RPCs. Three gas mixtures were tested which used the
same components, Ar, i-But and TFE, in different proportions:
TFE/Ar/i-But = 45/45/10; 60/27/13 and 75/15/10. In the three cases
the ratio Ar/TFE was changed to a large extent, living the i-But
concentration relatively stable.
\par
TFE is an heavy gas with good quenching properties \cite{5}. An
increase of TFE concentration at expenses of the Ar concentration
should therefore increase the primary ionization thus compensating
for the 40$\%$ reduction caused by the lower gas target pressure
(600 mbar) and reduce the afterpulse probability. For each of the
three gases a voltage scan was made for RPC2, leaving the other
two RPCs at a fixed operating voltage, and the following
measurements were made: RPC2 counting rate and current, double and
triple coincidence rate.
\par
\begin{figure}
  \begin{center}
    \leavevmode
    \includegraphics[width=10cm]{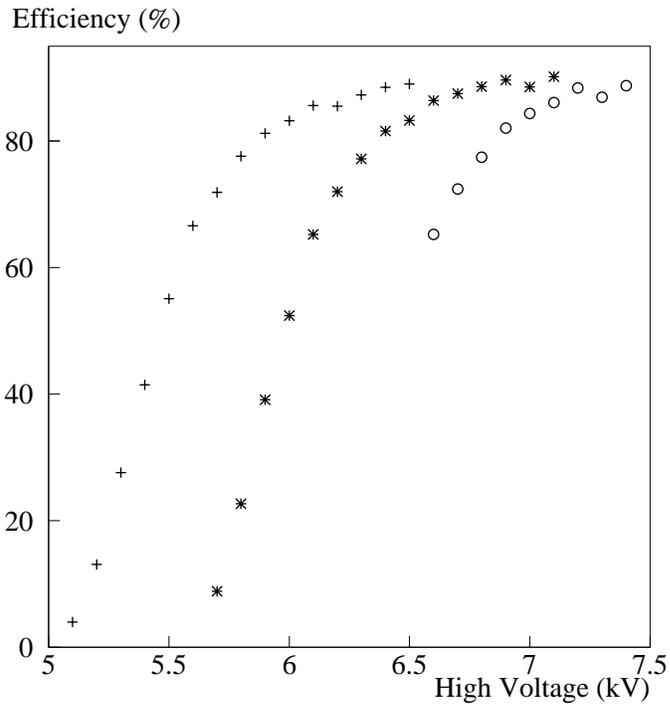}
    \caption{Detection efficiency of one RPC of the auxiliary
    telescope vs operating voltage for 3 gases: TFE/Ar/i-But=45/45/10
    (+); 60/27/13 (*) and 75/15/10 ($\circ$)}
    \label{3}
  \end{center}
\end{figure}
 The detection efficiency $vs$
the operating voltage for the three gases is shown in fig. \ref{3}
The reduction of the Argon concentration in favor of TFE results
in a clear increase of the operating voltage as expected from the
large quenching action of TFE. The data shown in fig. \ref {3} are
consistent with an increase of 30-40 V in operating voltage for a
1$\%$ reduction of the Argon concentration in the mixture.  In
spite of the different operating voltages all three gases approach
the same efficiency of about 90$\%$ which include the inefficiency
due to geometrical effects. A more systematic study of the plateau
efficiency is presented below, in connection with the carpet test.

\par
\begin{figure}[h!]
  \begin{center}
    \leavevmode
    \includegraphics[width=9cm]{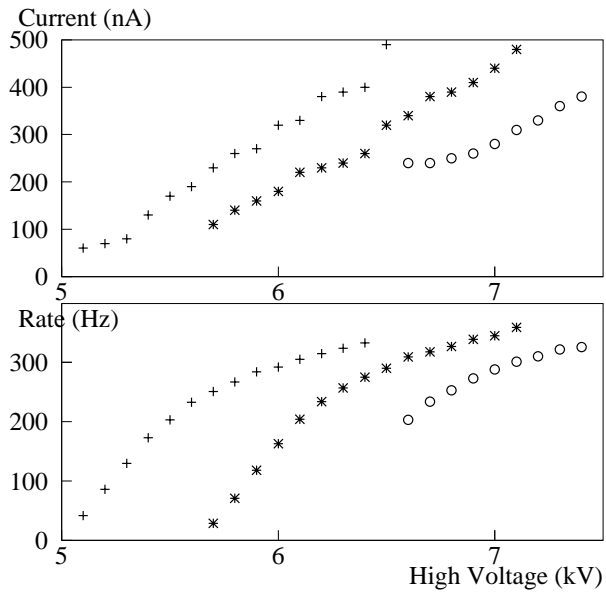}
    \caption{RPC2 operating current and counting rate vs voltage
for the three gases already mentioned in fig 3:
TFE/Ar/i-But=45/45/10
    (+); 60/27/13 (*) and 75/15/10 ($\circ$)}
    \label{4}
  \end{center}
\end{figure}

\begin{figure}[h!]
  \begin{center}
    \leavevmode
    \includegraphics[width=9cm]{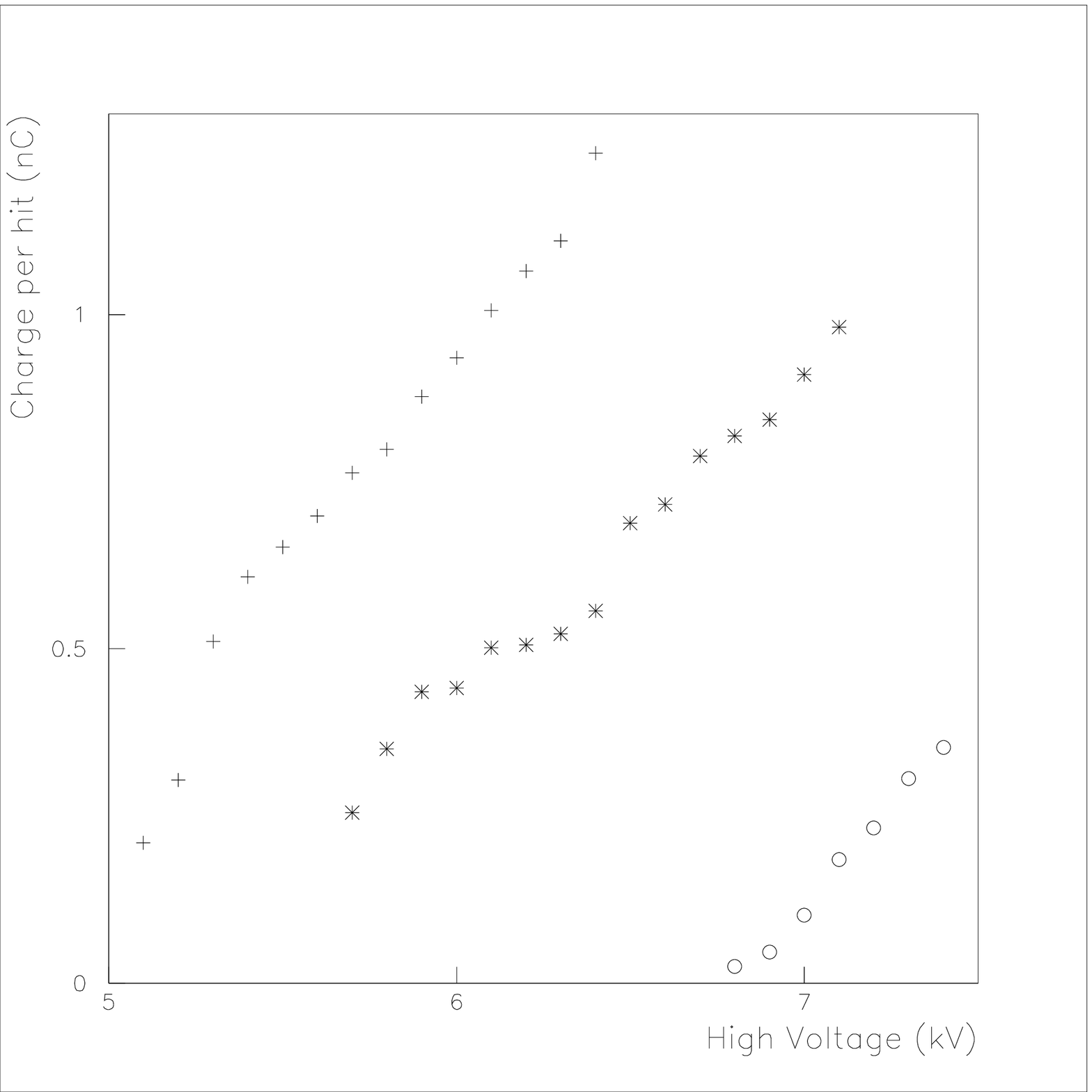}
    \caption{Charge delivered per count for the 3 gases vs operating voltage:
    TFE/Ar/i-But=45/45/10
    (+); 60/27/13 (*) and 75/15/10 ($\circ$)}
    \label{5}
  \end{center}
\end{figure}

Fig. \ref{4} shows the RPC2 current and counting rate vs the operating
voltage for the three gases. A small current linearly increasing
with the voltage is measurable well below the point where the RPC
start to show a significant counting rate. We interpret this as a
leak current not flowing through the RPC gas and not taking part
in the detector working mechanism.
\par
  The ratio of the operating current to the
counting rate gives the charge per count delivered in the RPC gas,
which is shown in fig. \ref{5} as a function of the operating
voltage for the three gases. Here the small term corresponding to
the current leaks, as mentioned above, is subtracted to the total
current. The data presented in fig. \ref{5} show that the higher
is the TFE fraction, the lower is the charge delivered in the gas
by a single streamer. Concerning the optimization of this
parameter the following points should be noted.
\begin{itemize}
  \item The signal
charge, in streamer mode operation, is anyway much above the
  achievable threshold of the front end electronics. This is particularly
  true for the final front end electronics that will be used for the
  experiment. Therefore a larger detector signal is not an advantage in
  this respect.
  \item A lower operating current, on the contrary, is an advantage
even if in a
  cosmic ray experiment the currents are expected to be modest.
  \item In a cosmic ray experiment, on the other hand, the analog
measurement of
  the hit density, which is achievable either from amplitude measurements
  of the strip signals or by sampling the operating current in appropriate
  time intervals, is an interesting possibility to be exploited for studing
  the shower core at energies as high as about 100 TeV.  Indeed, according
  to a MonteCarlo simulation of the final experiment, the digital read out
  of pads near the shower core, is expected to saturate at about 15-20 TeV.
  In this respect a lower delivered charge extends the dynamic range
  achievable for the analog measurement.
\end{itemize}
\par
\begin{figure}[h!]
  \begin{center}
    \leavevmode
    \includegraphics[width=10cm]{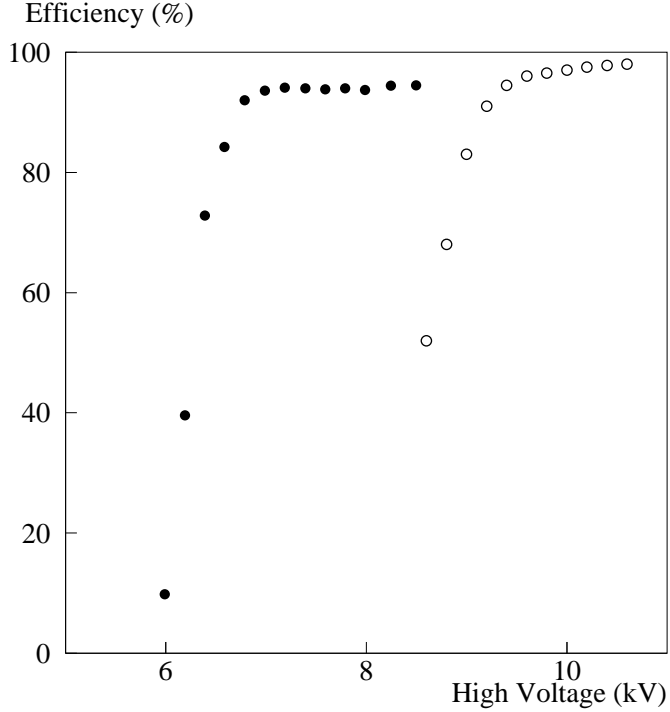}
    \caption{Detection efficiency vs operating voltage for one of the carpet
RPCs ($\bullet$). The same curve for a 2 mm gap RPC operating at
sea level is also reported ($\circ$) for comparison }
    \label{6}
  \end{center}
\end{figure}
We decided therefore to operate the test carpet with the gas
mixture  corresponding to the highest fraction of TFE.
\par
\begin{figure}[h!]
  \begin{center}
    \leavevmode
    \includegraphics[width=10cm]{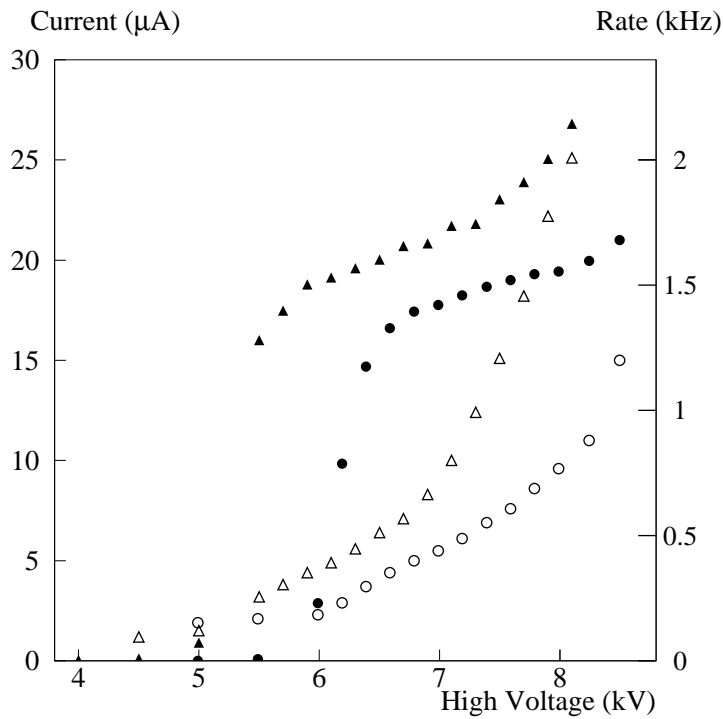}
    \caption{Counting rate (full triangles and circles) and operating current (open
             triangles and circles) vs Voltage of one RPC of the carpet. Results are presented for
             the gas with 45 $\%$ (triangles) and 75$\%$ (circles) of TFE respectively.
              The rate shown refers to 4 ORed Pads out of the 10 pads of
              the RPC. }
    \label{7}
  \end{center}
\end{figure}

The tests performed on the carpet were essentially the same as for
the auxiliary telescope. Fig. \ref{6} shows the operating
efficiency for the ORed pads 2-3-7-8 of one RPC of the carpet. The
efficiency was measured using cosmic ray signals defined by the
triple coincidence of the RPCs of the auxiliary telescope which
was placed on top of the carpet and centered on the corner among
four pads. The counting rate of the same pads OR signal, together
with the RPC current, are reported in fig. \ref {7} vs the
operating voltage. The results of the gas with TFE=45$\%$ are also
reported for comparison.  A rather flat counting rate plateau is
observed corresponding to a rate of about 400 Hz for a single pad
of area 56$\times$56 $cm^{2}$. The residual slope of the plateau
is mostly due to afterpulses occurring after the end of the 250 ns
shaped discriminated signals which produce a double counting of
the signal due to the same CR track. The rate and efficiency
curves rise in the same voltage interval as expected.
\par
\begin{figure}
  \begin{center}
    \leavevmode
    \includegraphics[width=10cm]{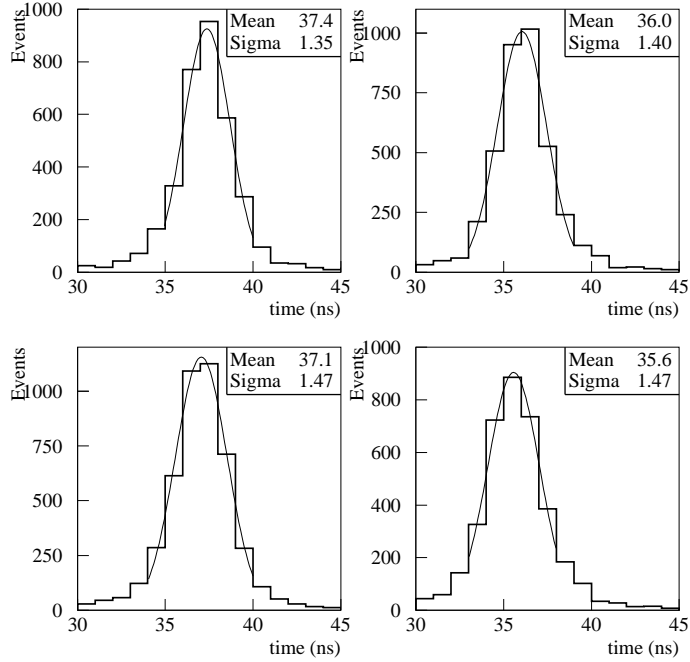}
    \caption{Time jitter distribution of 4 pads of the carpet.
The telescope RPC2 signal is used as common stop. The operating
voltage is 7.4 KV }
    \label{8}
  \end{center}
\end{figure}

The time jitter distribution of the pad signals was obtained by
measuring the delay of the fast OR signal with respect to RPC2 in
the trigger telescope.  This distribution is shown in fig. \ref{8}
for the four pads. The average of the standard deviations is 1.42
ns corresponding to a resolution of 1.0 ns for the single RPC if
we account for the fact that the distributions in fig. \ref{8}
show the combined jitter of two detectors.
\par
\begin{figure}
  \begin{center}
    \leavevmode
    \includegraphics[width=10cm]{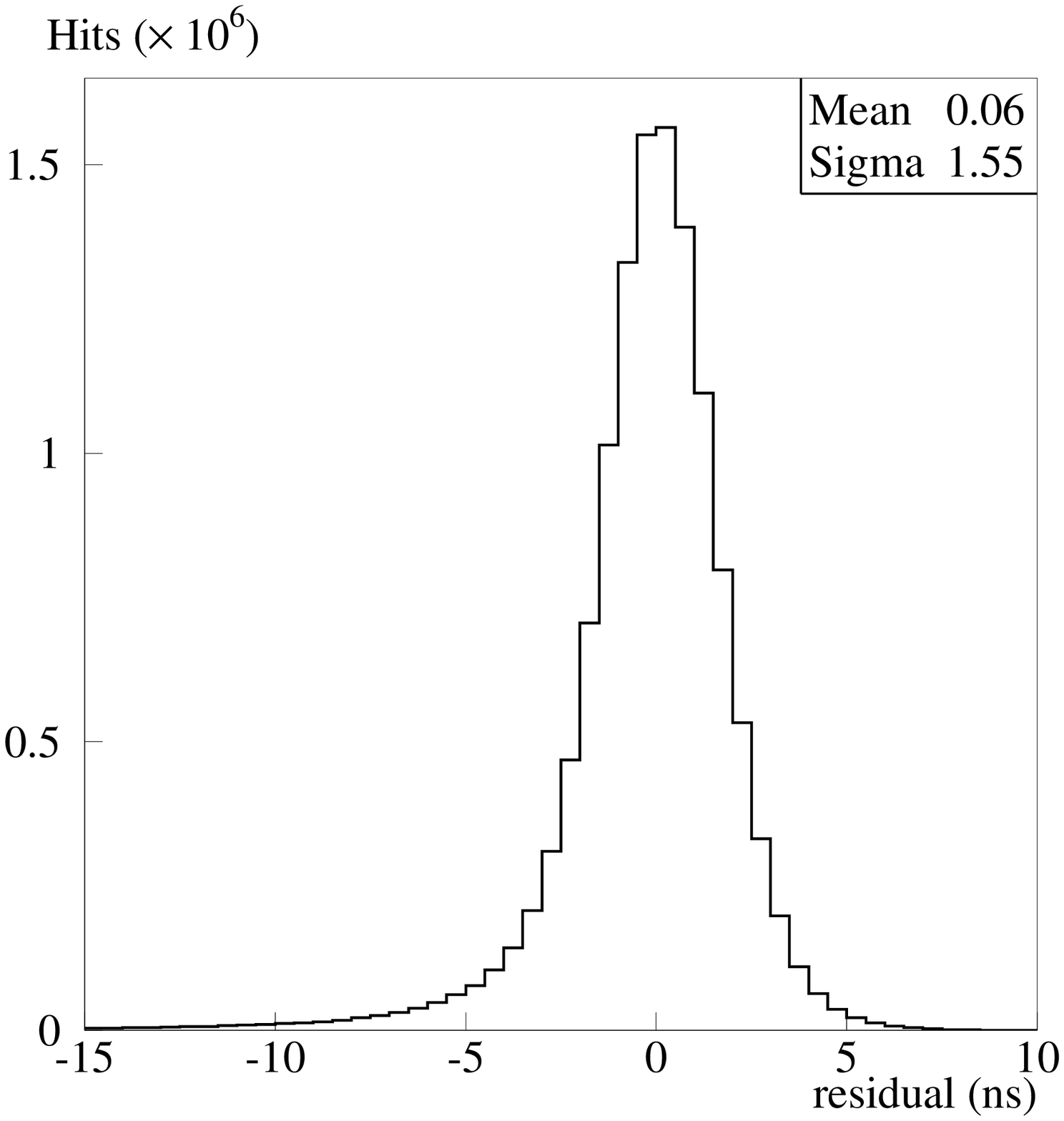}
    \caption{Time residual distribution for all 150 Pads. The trigger
signal is used as common stop}
    \label{9}
  \end{center}
\end{figure}

In the detection of extensive air showers however, the primary
cosmic ray direction is measured by comparing the times of hits
due to different particles of the shower. The space-time
distribution of the shower hits allows to fit the front of the
shower that can be assumed to a good approximation to be a plane.
The time residual distribution of the individual shower particles
with respect to the front is reported in fig. \ref{9}. The long
tail of delayed hits is due to particles arriving much after the
shower front.
\par

\section  { \bf{Discussion of the results}}

The use of RPCs for the detection of Extensive Air Showers in high
altitude laboratories poses some basic questions that the present
test contributes to answer:

\begin{itemize}
  \item  how do the operating voltage and plateau efficiency scale with
the pressure for the streamer mode operation

  \item how does the detector time resolution compare with the
intrinsic jitter of the shower front.
\end{itemize}

 With the purpose of
answering the first question a 2 mm gap RPC was operated at sea
level with the same gas, TFE/Ar/i-But=75/15/10, used for the YBJ
carpet. The detection efficiency vs operating voltage in fig.
\ref{6}, compared to the operation at 600 mbar pressure in YBJ,
shows an increase of about 2.5 kV in operating voltage.
\par
The effect
of small changes of temperature T and pressure P on the operating
voltage can be accounted for \cite{7} by rescaling the applied
voltage Va according to the relationship
$$
  V=V_{a}\frac{P_{0}}{P}\cdot \frac{T}{T_{0}}
$$ where Po and To are arbitrary standard values, e.g.1010 mb and
293 K respectively for a sea level laboratory.  However,  starting
from the YBJ data, the above formula predicts an operating voltage
at sea level which is considerably larger than the experimental
one.
\par
\begin{figure}
  \begin{center}
    \leavevmode
    \includegraphics[width=11cm]{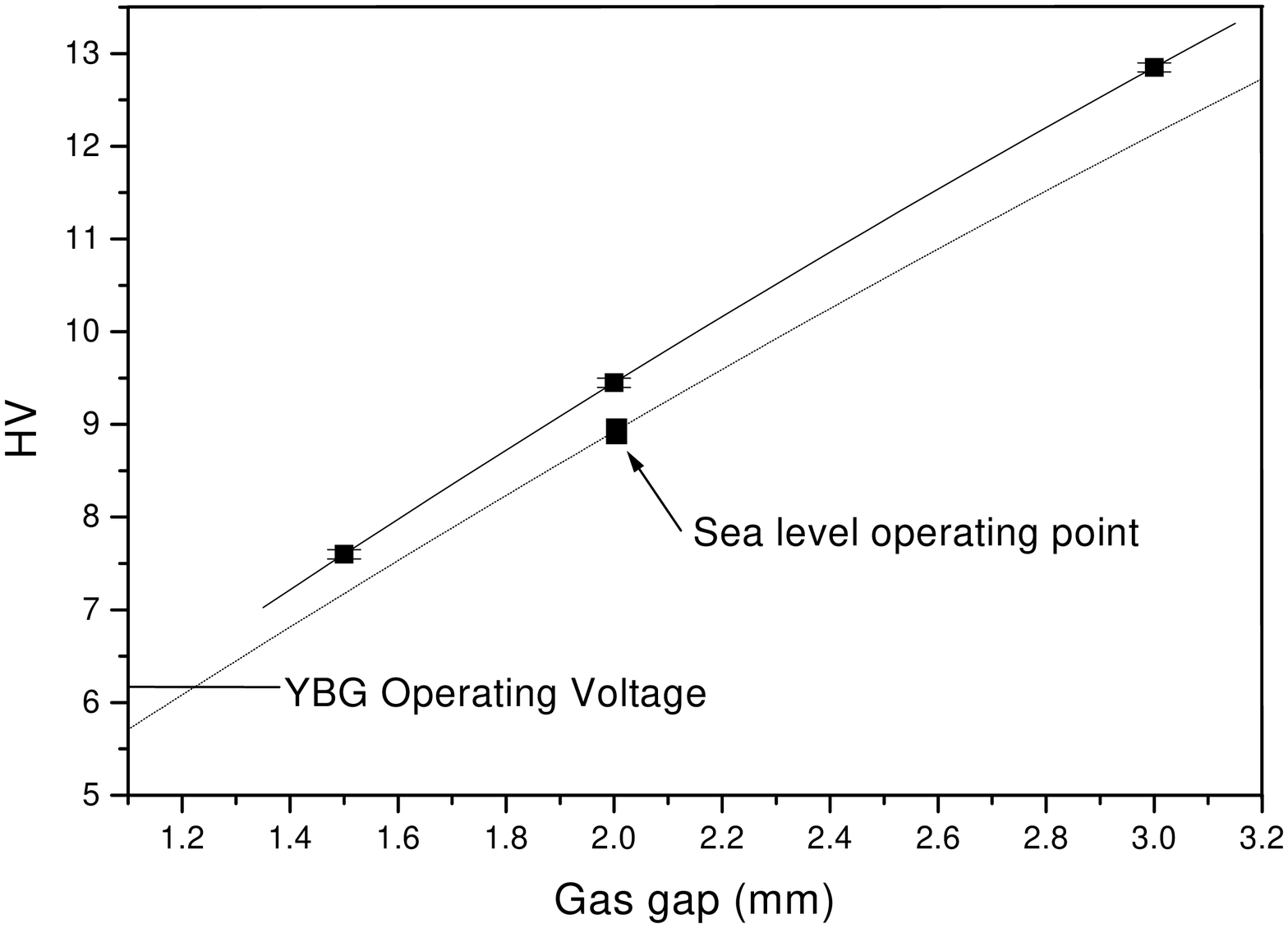}
    \caption{Operating voltage vs gas gap for the binary gas
TFE/i-But=97/3 (upper curve) and for the YBJ gas TFE/Ar/i-But=
75/15/10 (lower curve) }
    \label{10}
  \end{center}
\end{figure}

A large change of pressure produces a proportional change in the
gaseous target mass per unit surface, like a change of the gas gap
size. The operating voltage as a function of the gap is studied in
\cite{8} for 1.5, 2, and 3 mm gap RPCs, in the case of the binary
gas mixture TFE/i-But=97/3. The result is shown in fig. \ref{10}
where the operating voltage in streamer mode is defined as that
giving 50$\%$ streamer probability with respect to the plateau.

The data which refer to the same pressure of 1010 mbar and
temperature of 293 K, show that the voltage do not scale
proportionally to the gap, the electric field (voltage/gap) being
larger for thinner gaps. Indeed the avalanche to streamer
transition occurs when the gas amplification, $e^{\alpha g
}/\alpha g$, exceeds a given threshold. The larger is the gap the
smaller is the $\alpha$ value and therefore the electric field
that is needed for reaching the streamer threshold. A zero
constraint parabolic fit of the three experimental points is also
reported in fig. \ref{10}. The fitted curve, which refers to the
binary gas, can be scaled to the YBJ gas, TFE/Ar/i-But=75/15/10 at
$20 ^{O}C$ temperature, using the point at sea level (8.6 kV at
1010 mbar and $32 ^{O}C$, rescaled to $20 ^{O}C$ according to the
above formula) and assuming that the ratio of the operating
voltages for the two gases is the same for all gap sizes.  The
result is the lower curve in fig. \ref{10} which represents the
operating voltage vs gap for the YBJ gas and fits well the YBJ
operating point, 6.12 kV, if we assume that a 2mm gap at the YBJ
pressure of 603 mbar is equivalent to a 1.2 mm gap at 1010 mbar.
The above assumption is based on the fact that, in the ideal gas
approximation, the mass per unit surface of the gaseous target,
which fixes the operating voltage for each gas, is given by the
parameter $gap \cdot pressure / temperature$.

Fig. \ref{6} also shows that the plateau efficiency measured at
YBJ is 3-4$\%$ lower than at sea level. Although a lower
efficiency is expected from the smaller number of primary clusters
at the YBJ pressure, we attribute most of the difference to the
underestimation of the YBJ efficiency. At the YBJ level indeed the
ratio of the cosmic radiation electromagnetic to muon component is
about 4 times larger than at sea level. A spatial tracking with
redefinition of the track downstream of the carpet would eliminate
the contamination from soft particles, giving  a more accurate and
higher efficiency. On the other hand the lower efficiency could
hardly be explained with the gas lower density. The number of
primary clusters in the YBJ test, estimated around 9, is the same
as in the case of some gas, e.g. Ar/iBut/CF3Br=60/37/3, that was
frequently used at sea level with efficiency of  97-98$\%$
\cite{9}.
\par
The time residual distribution in fig. \ref{9} shows a long tail
due to delayed particles traveling well behind the shower front.
The gaussian fit, disregarding this tail, gives a standard
deviation of 1.6 ns to be compared with the RPC intrinsic time
resolution of 1.0 ns. Taking into account the additional
uncertainties due to the propagation time of the signal traveling
along a strip of 56 cm and to the impact point of the shower
particle which can be everywhere inside the PAD we get a total RPC
jitter of 1.3 ns. The residual jitter of the shower front can be
estimated to be: $ \sigma_{shower} = 0.9 ns. $

This is valid for high energy showers selected by the multiplicity
trigger as in the case reported in fig. \ref{9} At lower energies
the shower jitter increases gradually.

\begin{ack}
The authors are endebted to G. Aielli (Universit\`{a} di Roma ``Tor Vergata'') for
editing the present paper.
\end{ack}

\listoffigures
\end{document}